\documentstyle[12pt]{article}
\begin{document}
\begin{center} {\Large {\bf 
Nonequilibrium Multicritical Behaviour in Anisotropic
Heisenberg Ferromagnet driven by Oscillating Magnetic Field }} \end{center}

\vskip 1 cm

\begin{center}{Muktish Acharyya}\end{center}
\begin{center}{\it Department of Physics, Krishnanagar Government College,
PO-Krishnanagar, Dist-Nadia, PIN-741101, WB, India}\end{center}
\begin{center}{\it E-mail: muktish@vsnl.net}\end{center}
\vskip 2cm

The Heisenberg ferromagnet (uniaxially anisotropic along z-direction), in the presence of time dependent (but uniform over space)
magnetic field, is studied by Monte Carlo simulation. The time dependent magnetic field was taken as elliptically
polarised in such a way that the resulting field vector rotates in the XZ-plane. In the limit of low anisotropy, the
dynamical responses of the system are studied as functions of temperature and the amplitudes of the magnetic field.
As the temperature decreases, it was found that the system undergoes multiple
dynamical phase transitions. In this limit, the multiple transitions 
were studied in details and the phase
diagram for this observed multicritical behavior was drawn in field amplitude and temperature plane. 
The natures (continuous/discontinuous)
of the transitions are determined by the temperature variations of fourth-order Binder cumulant ratio and the distributions
of order parameters near the transition points.
The transitions are supported by finite size study. The temperature variations of the variances of dynamic order
parameter components (for different system sizes) indicate the existence of diverging length scale near the dynamic
transition points. The frequency dependences of the transition temperatures
of the multiple dynamic transition are also studied briefly.

\vskip 0.5cm

\begin{center} {\bf INTRODUCTION} \end{center}

The nonequilibrium dynamical phase transition \cite{rmp}, particularly in the kinetic Ising model, has drawn immense
interest of the researchers working in the field of nonequilibrium statistical physics. The dynamic transition in the
kinetic Ising model in the presence of sinusoidally oscillating magnetic field was noticed \cite{tome} first in the
mean field solution of the dynamical equation for the average magnetisation. Due to the absence of fluctuations, in
the mean field study, the dynamic transition would be possible also even in the static (zero frequency) limit. But the
true dynamic transition should never occur in the static limit due to the presence of nontrivial 
thermodynamic fluctuations. The occurence of the
true dynamic transitions for models, incorporating the thermodynamic fluctuations, was later shown in several Monte
Carlo studies \cite{rmp}.

Since, the Ising model is quite simple and a prototype to study the nonequilibrium 
dynamical phase transitions, a considerable amount
of work was done on it \cite{rmp}. Among these, several studies were done to establish that all these transitions 
have some features which are similar to
well known thermodynamic phase transitions. The divergence of "time scale" \cite{ma1} and the "dynamic specific heat"
\cite{ma1} and the divergence of "length scale" \cite{rikpre} are two very important observations to establish that the
dynamic transition, in kinetic Ising model driven by sinusoidally oscillating magnetic field, 
is indeed a thermodynamic phase transition. 

Although, the dynamic phase transition in kinetic Ising model is an interesting phenomenon which acts as a simple example to grasp the
various features of nonequilibrium phase transitions, it has several limitations. Since the orientations of the spins
are limited by only two (up/down) directions, some interesting features of dynamic transitions (related to the dynamical
transverse ordering) cannot be observed in Ising model. The classical vector spin model \cite{matis} would be 
the proper choice to study such interesting phenomena which are missing in Ising model. The "off-axial" dynamic transition 
\cite{ijmpc}, recently observed in Heisenberg ferromagnet, is an example of such interesting phenomena. In the "off-axial"
dynamic transition, the dynamical symmetry along the axis of anisotropy can be broken dynamically by applying an 
oscillating magnetic field along any perpendicular direction. There are several important studies done on classical
vector spin models driven by oscillating magnetic field. The dynamic phase transition in an anisotropic XY ferromagnet
driven by oscillating magnetic field was recently studied \cite{yasui} by solving Ginzburg-Landau equation. The dynamic 
phase transition and the dependence of its behaviour on the bilinear exchange anisotropy of classical Heisenberg ferromagnet
(thin film), was studied by \cite{jang} Monte Carlo Simulation. The dynamic transitions alongwith hysteresis scaling 
in Heisenberg ferromagnet was also studied \cite{huang} both by Monte Carlo simulation and mean field solution. 

All these
studies (discussed in above paragraph) of dynamic transitions in classical vector spin model give single dynamic transitions. There
are a few studies in Heisenberg ferromagnets where the multiple dynamic transition was reported \cite{mdt1,mdt2}. A recent
study \cite{mdt1} of dynamical phase transition in thin Heisenberg ferromagnetic films with bilinear exchange
anisotropy has shown multiple phase transitions for the surface and bulk layers of the film at different temperatures.
However, another study \cite{mdt2} shows triple dynamic transitions (at three different temperatures), for bulk dynamic ordering only,
in uniaxially anisotropic Heisenberg ferromagnet driven by elliptically polarised magnetic field. In the present paper, the multiple
dynamic transition (bulk only) \cite{mdt2}
in uniaxially anisotropic Heisenberg ferromagnets driven by elliptically polarised magnetic field, is studied extensively. 

In this paper, the multiple dynamic transition, in the uniaxially anisotropic classical Heisenberg ferromagnet
in presence of elliptically polarised magnetic field, is studied by Monte
Carlo simulation. The various dynamical phases were found. The multiple dynamic phase transition was observed by studying the
temperature variation of the dynamic specific heat and the derivatives of dynamic order parameter components. 
The nature (continuous/discontinuous) of the transition was detected by studying the distribution of order parameter and the
temperature variation of Binder cumulant ratio near the transition point.
A finite size
study was done supporting the multiple dynamic phase transitions. A length scale was found to diverge near the dynamic transition
points. And finally and most importantly, the phase diagram for this multiple dynamic phase transition was drawn. The frequency
dependence of these multiple dynamic transitions was also studied briefly.

The paper is organized as follows: in the next section the description of the system, i.e., the Heisenberg ferromagnet in
polarised magnetic field, is given, section III describes the Monte Carlo simulation technique employed here to study
the dynamical steady state of this type of classical
vector spin model, in detail. The numerical results with diagrams are reported in section IV and the paper ends with a 
summary in section V.
\vskip 1 cm

\begin{center} {\bf HEISENBERG FERROMAGNET IN POLARIZED MAGNETIC FIELD} \end{center}

The classical anisotropic (uniaxial and single-site) Heisenberg model, with nearest neighbour ferromagnetic interaction
in the presence of a magnetic field can be described by the following Hamiltonian:
\begin{equation}
H = -J\Sigma_{<ij>} {\bf S_i} \cdot {\bf S_j} - D \Sigma_i (S_{iz})^2 - {\bf h} \cdot \Sigma_i {\bf S_i}.
\label{hamiltonian}
\end{equation}
\noindent In the above expression, ${\bf S_i}[S_{ix},S_{iy},S_{iz}]$ represents a classical spin vector
(situated at the $i$th lattice site) of magnitude
unity, i.e., $|{\bf S}|=1$ or $S^2_{ix}+S^2_{iy}+S^2_{iz} = 1$. The classical spin vector ${\bf S_i}$ may take any (unrestricted) 
angular orientation in the vector spin space. The first term, in the Hamiltonian, represents the nearest -neighbour ($<ij>$)
ferromagnetic ($J>0$) interaction. The factor $D> 0$, in the second term, represents the strength of uniaxial ($z$ axis here)
anisotropy which is favoring the spin vector to be aligned along the $z$ axis. This is readily seen since the second term
minimizes energy for maximization of the value of $S^z_i$. Here, it may be noted that for $D \rightarrow \infty$ the system
goes to the Ising limit and for $D=0$ the system is in the isotropic Heisenberg limit. The last term stands for the interaction with
the externally applied time dependent magnetic field $[{\bf h}(h_x,h_y,h_z)]$. The magnetic-field components are sinusoidally
oscillating in time, i.e., $h_{\alpha} = h_{0\alpha}{\rm cos}(\omega t)$, where $h_{0\alpha}$ is the amplitude and $\omega$ is
the angular frequency ($\omega = 2 \pi f; f$ is the frequency) of the $\alpha$th component of the magnetic field. In the present
study, the field is taken elliptically polarised. In general, the field vector is represented as
\begin{equation}
{\bf h} = {\hat x}h_x + {\hat y}h_y + {\hat z}h_z \\
= {\hat x}h_{0x}{\rm cos}(\omega t) + {\hat z} h_{0z} {\rm sin} (\omega t).
\end{equation}

\noindent The time eliminated relation between $h_x (= h_{0x} {\rm cos} (\omega t))$ and $h_z (=h_{0z} {\rm sin} (\omega t)$
is
\begin{equation}
{h_x^2 \over h_{0x}^2} + {h_z^2 \over h_{0z}^2} = 1.
\end{equation}

\noindent This is the equation of an ellipse which indicates that the magnetic field vector is 
elliptically polarised (for $h_{0x} \neq h_{0z}$) and lies on the $X-Z$ plane. For $h_{0x}=h_{0z}=h_0$ (say), the field 
will be circularly polarised and $h_x^2 +h_z^2 = h_0^2$. The magnetic fields and the strength of anisotropy $D$ are
measured in units of $J$. The model is defined on a simple cubic lattice of linear size $L$ with periodic boundary
conditions applied in all three ($x, y, z$) directions.

\vskip 1 cm

\begin{center} {\bf MONTE CARLO SIMULATION METHOD} \end{center}

Monte Carlo (MC) simulation method was employed to study the above described model. The algorithm is \cite{stauffer} described below.
The system is slowly cooled down from a random initial spin configuration \cite{uli}, to obtain the steady state spin configuration
at a particular temperature $T$ (measured in units of $J/k_B$, where $k_B$ is the Boltzmann constant).
The initial random spin configuration was generated as follows \cite{ijmpc,uli}: two different
(uncorrelated) random numbers $r_1$ and $r_2$ (uniformly distributed between -1 and +1), are chosen in such a way that
$R^2 = r_1^2 + r_2^2$ becomes less than or equal to unity. The set of values of $r_1$ and $r_2$, for which $R^2 > 1$, are rejected.
Now, $S_{ix} = 2ur_1$, $S_{iy} = 2ur_2$ and $S_{iz} = 1- 2R^2$, where $u = \sqrt{1-R^2}$. After preparing initial random configuration
of spins (this is the proper choice of the spin configurations,
 corresponding to very high temperature), one has to find a steady state configuration
for any fixed temperature $T$.
For any fixed set of values of $h_{0x}$, $h_{0z}$, $\omega$ and $D$ and at any particular temperature $T$, a lattice site $i$
has been chosen randomly (random updateing scheme). The value (random direction) of the spin vector at this randomly chosen site
is ${\bf S_i}$ (say). The energy of the system is given by the Hamiltonian (Eq. \ref{hamiltonian}). 
A test spin vector ${\bf S'_i}$ is then chosen (for a trial move)
at any random direction (following the same algorithm described above). For this choice of ${\bf S'_i}$, at site $i$ the energy will
be 
$H = -J\Sigma_{<ij>} {\bf S'_i} \cdot {\bf S_j} - D \Sigma_i (S'_{iz})^2 - {\bf h} \cdot \Sigma_i {\bf S'_i}$.
The change in energy, associated with this change in the direction of spin vector (from ${\bf S_i}$ to ${\bf S'_i}$ at lattice site 
$i$), is $\Delta H = H'-H$. The Monte Carlo method \cite{stauffer} will decide whether this trial move is acceptable or not. 
The probability of this move (chosen here) is given by Metropolis rate \cite{metro}
\begin{equation}
W({\bf S_i} \rightarrow {\bf S'_i}) = {\rm Min}{\Large[}1,{\rm exp}{\Large (}-{{\Delta H} \over {k_B T}}{\Large )}{\Large ]}.
\end{equation}
\noindent Now, this probability will be compared with a random number $R_p$ (say) (uniformly distributed between zero and one).
If $R_p$ does not exceed $W$, the move (${\bf S_i} \rightarrow {\bf S'_i}$) will be accepted. In this way the spin vector ${\bf S_i}$
is updated. $L^3$ numbers of such random updates of spins, defines one Monte Carlo step per site (MCSS) and this is considered as the unit
of time in this simulation. 
The efficiency of the MC technique can be increased by choosing the direction of the spin for the trial move close to the 
present direction which will increase the probability of acceptance.
So, 50 MCSS are required to have one complete cycle of the oscillating magnetic field (of frequency $f = 0.02$) and hence the time
period ($\tau$) of the field becomes 50 MCSS. Any time dependent macroscopic quantity (i.e., any component of magnetisation at any
instant $t$) is calculated as follows: Starting with an initial random spin configuration (corresponding 
to the high-temperature disordered phase),
the system is allowed to become stable (dynamically) up to $5\times10^4$ MCSS (i.e., 1000 complete cycle of the oscillating field). 
The average value of various physical quantities are calculated from further $5\times10^4$ MCSS (i.e., averaged over another 1000
cycles). This was checked carefully that the number of MCSS mentioned above is sufficient to achieve dynamical 
steady state value of the
measurable quantities, etc. which would clearly show the dynamic transition points within limited accuracy. But to describe the
critical behaviour very precisely (i.e., to estimate critical exponents etc) a much longer run is required. The total length of the
simulation becomes $10^5$ MCSS. The system is slowly cooled down ($T$ has been reduced by a small interval 
$\delta T = 0.02$ here) to get the values
of the statistical quantities in the lower temperature phase. 
Here, the last spin configuration corresponding to the previous
temperature is employed to act as initial configuration for the new (lower) temperature. The CPU time required for $10^5$ MCSS is
approximately 30 min on an Intel Pentium-III processor.

One important point may be noted here regarding the dynamics chosen in this simulational study. Since the spin components do not
commute with the Heisenberg Hamiltonian, it has intrinsic 
quantum mechanical dynamics. Considering the intrinsic dynamics, there was a study \cite{krech}
of the structure factor and transport properties in XY model. However, the present paper aims to study the nonequilibrium phase transition
governed by thermal fluctuations only. Keeping this in mind, one should choose the dynamics that arising solely only due to 
the interaction with thermal
bath. Since the objective is different, in this paper, the dynamics chosen (arise solely due to the interaction with a thermal
bath) were Metropolis dynamics. The effect of intrinsic spin dynamics is therefore not considered here.
 
\vskip 1 cm

\begin{center} {\bf NUMERICAL RESULTS} \end{center}

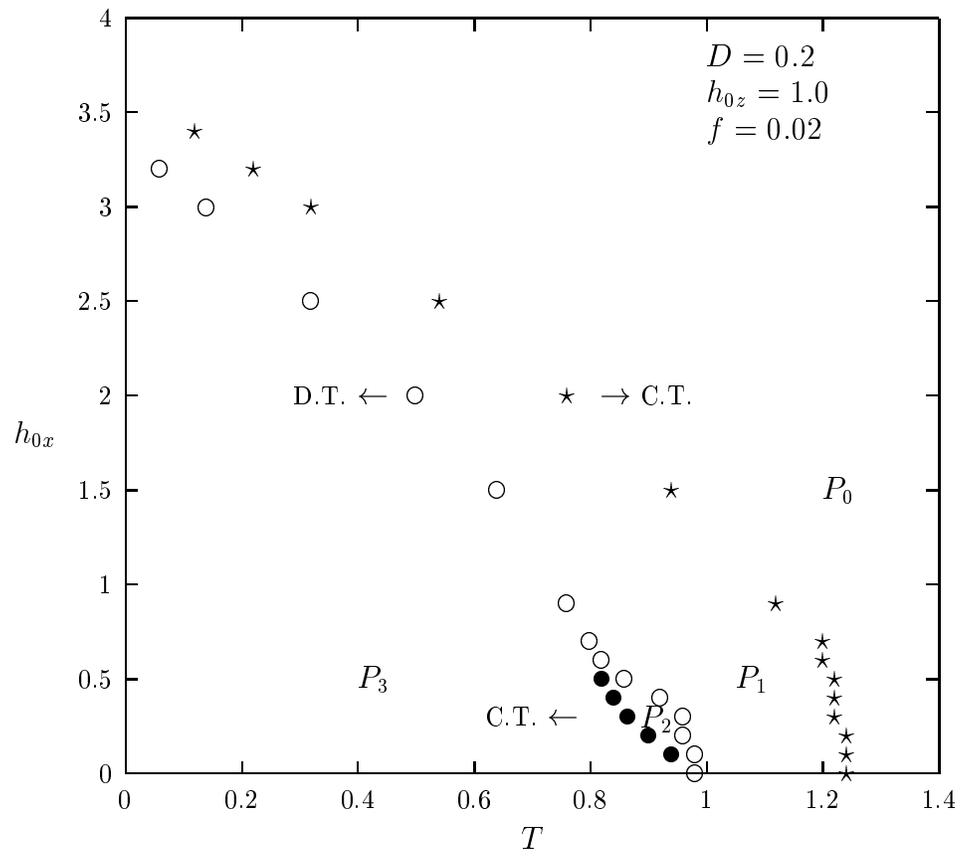
\begin{figure}
\caption {The phase diagram in $T-h_{0x}$ plane for the multiple dynamic transitions for $D=0.2$ 
$f = 0.02$ and $h_{0z} =1.0$. C.T. stands for continuous transition and D.T. stands for discontinuous
transition. The natures (continuous/discontinuous) of the transitions across the different phase boundary
are marked by C.T and D.T.}
\label{phase}
\end{figure}

The Monte Carlo simulations, in the present study, were done on a simple cubic lattice of linear size $L = 20$. 
For a fixed set of values of amplitudes ($h_{0x}, h_{0z}$) and frequency ($f$) of 
polarised magnetic field, strength of anisotropy ($D$) and temperature ($T$), the instantaneous
magnetisation components (per lattice site) were calculated as follows: $m_x(t)=\Sigma_i S^x_i/L^3 $, $m_y(t)=\Sigma_i S^y_i/L^3 $
and $m_z(t)=\Sigma_i S^z_i/L^3 $. The dynamic order parameter is defined as the time averaged magnetisation over a full
cycle of the oscillating magnetic field. In this case the dynamic order parameter ${\bf Q}$ is a vector (since the
magnetisation is vector). The components of the dynamic order parameter are calculated as $Q_x=(1/\tau)\oint m_x(t) dt$,
$Q_y=(1/\tau)\oint m_y(t) dt$ and $Q_z=(1/\tau)\oint m_z(t) dt$. 
Here, four different dynamical phases are identified. The high temperature disordered phase is denoted as
$P_0 : (Q_x =  Q_y = Q_z = 0)$. Other three ordered phases are $P_1 : (Q_x = 0, Q_y \neq 0, Q_z =0)$, $P_2 :
(Q_x \neq 0, Q_y = 0, Q_z \neq 0)$ and $P_3 : (Q_x = 0, Q_y = 0, Q_z \neq 0)$. The phase diagram, obtained from multiple
dynamic transitions, is plotted in $h_{0x} - T$ plane and is shown in fig \ref{phase}. The methods of getting the phase
boundaries are discussed in the following paragraphs.

The signature of such successive phase transitions are also observed by studying the temperature variations of
the magnitude of the
dynamic order parameter, energy, specific heat and the temperature derivatives of dynamic order parameter components.
For the same set of values $D=0.2$, $h_{0x} = 0.3$ and $h_{0z} =1.0$ the temperature variations of various dynamic
quantities are studied and shown in Fig. \ref{pt3}. The temperature variations of the dynamic order parameter components
$Q_x$, $Q_y$ and $Q_z$ are shown in Fig. \ref{pt3}(a). As the system is cooled down, from a high-temperature dynamically
disordered (${\bf Q} = \vec 0$) phase, it was observed that first the system undergoes a transition from dynamically disordered
$P_0$ phase 
(${\bf Q} = \vec 0$) to a dynamically Y-ordered ($Q_y \neq 0$ only) phase $P_1$ and the transition temperature is $T_{c1}$ (say).
It may be noted here that the resultant vector of elliptically polarised magnetic field lies in the $x-z$ plane and the dynamic
ordering occurs along the $y$-direction only ($Q_x =0$, $Q_y \neq 0$, $Q_z = 0$). This is 
clearly an off-axial transition \cite{ijmpc}. In the case
of this type of off-axial transition the dynamical symmetry (in any direction; $y$ here) is broken by the application of the
magnetic field in the perpendicular (lies in the $x-z$ plane here) direction. As the system cools down further, this phase $P_1$
persists over a considerable range of temperature and at a temperature $T_{c2}$ a second transition was observed. In this phase, the
system becomes dynamically ordered both in the $X$ and $Z$ directions at the cost of $Y$ ordering. Usually one gets
the dynamically ordered phase of second kind $P_2: (Q_x \neq 0,
Q_y = 0, Q_z \neq 0)$. Here, the dynamical ordering is planar (lies on $x-z$ plane) and the dynamical ordering occurs in the same
plane on which the field vector lies. This transition is not of off-axial type. As the temperature decreases further one ends up with
a low-temperature dynamically ordered phase of third kind 
$P_3 : (Q_x = 0, Q _y =0, Q_z \neq 0)$ via a transition occurs at temperature $T_{c3}$. 
The system continues to increase the dynamical $Z$ ordering ($Q_z \neq 0$ only)
as the temperature decreases further. The low-temperature phase is only dynamically $Z$ ordered. No further transition was 
observed as one cools the system further. From Fig.\ref{pt3}(a), it is quite clear, if one observes carefully, that the
sizes of the errorbars of $Q_x$, $Q_y$ and $Q_z$ are maximal near the transition points, indicating the growth of 
fluctuations near the transiton points. 

One gets qualitative ideas of multiple dynamic phase transitions from the above mentioned studies. However, to estimate precisely the
transition temperatures $T_{c1}$, $T_{c2}$ and $T_{c3}$ further studies are required. For this reason, the temperature variations
of the derivatives (with respect to temperature) of the dynamic order parameter components are studied. 
The derivatives were calculated numerically by using central difference formula \cite{mth}
\begin{equation}
{{df(x)} \over dx} = {{f(x+\delta x) - f(x-\delta x)} \over {2\delta x}}.
\label{derivative}
\end{equation}
Fig. \ref{pt3}(b) shows
such variations studied as a function of temperature. The derivative ${{dQ_y}/{dT}}$, shows a sharp minimum at 
$T=T_{c1} \simeq 1.22$. The second transition temperature $T_{c2}$ was estimated from the 
position of sharp maximum of ${{dQ_y}/{dT}}$ and the corresponding sharp minima of 
${{dQ_x}/{dT}}$ and ${{dQ_z}/{dT}}$. This gives $T_{c2} \simeq 0.96$. At slightly lower temperature 
than $T_{c2}$, one observes
another maximum of ${{dQ_x}/{dT}}$ and minimum of ${{dQ_z}/{dT}}$ both 
at the same position $T=T_{c3} \simeq 0.88$. From this study
one gets the quantitative measure of the transition temperatures of multiple dynamic transition \cite{mdt2}. 
The maximum error, in estimating the transition temperatures, is 0.01.

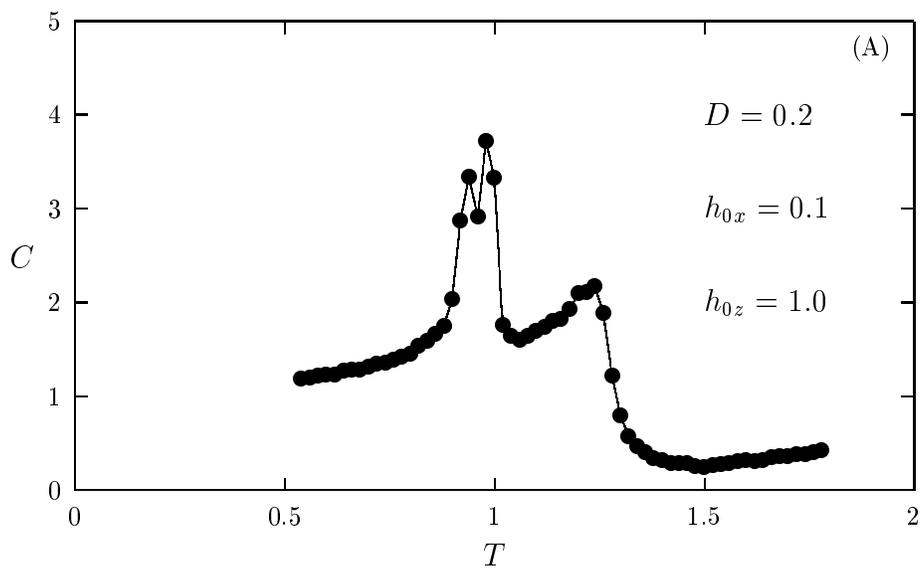
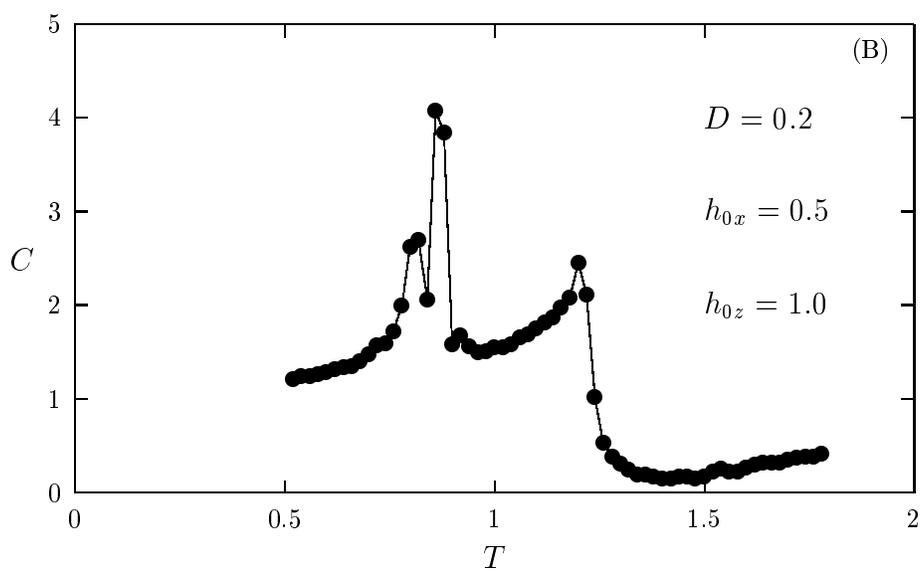
\begin{figure}
\caption {The temperature variations of dynamic specific heat for $f=0.02$, $D=0.2$, $h_{0z} =1.0$.
(A) for $h_{0x} = 0.1$ and (B) for $h_{0x} = 0.5$.
Solid lines are guide to the eye. The maximum error in estimating the transition temperature is 0.01.}
\label{oth2}
\end{figure}

Similar values of the transition temperatures ($T_{c1}$, $T_{c2}$, $T_{c3}$) for the multiple dynamic transition can be
estimated independently from the study of the temperature variations of dynamic energy and specific heat. The dynamical energy ($E$)
is defined as the time averaged value of the instantaneous energy over a full cycle of the oscillating magnetic field.
From the definition, $E = (1/\tau)\oint H dt$ ($H$ is given in Eqn. \ref{hamiltonian}). The temperature variation of $E$ is
studied and shown in Fig. \ref{pt3}(c). This shows two inflection points and one (the middle one) discontinuity 
at the same location of transition temperatures
estimated and mentioned above. This will be very clear if one studies the derivative of the dynamic energy, namely the dynamic
specific heat $C = {dE/dT}$ (calculated by using central difference formula \ref{derivative}). The temperature variation
of dynamic specific heat $C$ was studied and shown in Fig. \ref{pt3}(d). It indicates three peaks at three different temperatures
supporting $T_{c1} \simeq 1.22$, $T_{c2} \simeq 0.96$ and $T_{c3} \simeq 0.88$. 
Thus the estimation of transition temperatures for the multiple
dynamic transition was reexamined by another independent study. The study of the temperature variation of dynamic specific heat
has another importance. It independently supports multiple transitions as well as it indicates that these transitions
are indeed thermodynamic phase transitions. 
Now one may employ both the methods of studying the temperature 
variations of derivatives of dynamic order parameter components
and the dynamic specific heat to estimate the transition points.
One gets three different dynamic transitions for the parameter values $D = 0.2$,
$h_{0x} =0.3$ and $h_{0z} = 1.0$. 
This three transitions senario is observed for a range of values of $h_{0x}$ (keeping other parameters fixed $D=0.2$ and
$h_{0z}=1.0$) between $h_{0x} =0.1$ to $h_{0x} = 0.5$ (within the precision considered here).
The temperature variations of dynamic specific heat $C$ are shown in fig \ref{oth2} for two different values
of $h_{0x}$ (= 0.1 and 0.5). Both show the three dynamic transitions.
It may be noted from fig. \ref{oth2} that the transition temperature decreases as the amplitude $h_{0x}$ increases.
Now, let us see what happens if one takes the value of the $x$-amplitude of elliptically polarised magnetic
field, i.e., $h_{0x}$, outside this range, keeping all other parameter values unchanged. 

For $h_{0x} =0.0$ (effectively the field is now linearly polarised along $z$ direction only), to estimate the transition
points the temperature variations of dynamic specific heat was studied and plotted in fig. \ref{pt0}(a). This shows 
two distinct and well separated peaks indicating two  
dynamic transitions, one at $T \simeq 1.24$ and other at $T \simeq 0.98$. Now to characterise the different phases the temperature
variations of dynamic order parameter components were studied. This study is shown in fig. \ref{pt0}(b).
This shows that the system starts to get dynamically ordered from a high-temperature disordered phase.
So, the high-temperature {\it ordered phase}
 ($Q_x \neq 0$, $Q_y \neq 0$, $Q_z = 0$) is quite different from $P_1$ (described above for $h_{0x} \neq 0$). The second
(low-temperature ordered phase) is $P_3$ type. This transition (from high-temperature ordered phase to low-temperature
ordered phase) occurs at $T_{c3} \simeq 0.98$. So, one gets two distinct transitions for $D=0.2$, $h_{0x}=0.0$ and $h_{0z} =1.0$.
It may be noted here that the temperature variations of the order parameter components, mainly $Q_x$ and $Q_y$,
are quite scattered (near the low-temperature transition). 
Here, $Q_x$ and $Q_y$ are calculated by averaging over 2000 cycles discarding first 2000 cycles ($10^5 MCSS$).
For this reason the derivatives $dQ_x/dT$ and $dQ_y/dT$ do
not give smooth variation with respect to temperature and are not shown.

Now if the value of $h_{0x}$ is higher (say around $h_{0x} =0.6$) with all other parameter values fixed (i.e.,
$f=0.02$, $D=0.2$ and $h_{0z} = 1.0$) the triple dynamic transitions (observed for $h_{0x} =0.3$) reduces to double transitions.
It gives two distinct peaks of $C$ indicating two different dynamic transitions at $T \simeq 1.20$ and $T \simeq0.82$. 
So, in this case one gets 
two dynamically ordered phases, namely $P_1$ (higher temperature ordered phase) and $P_3$ (lower temperature ordered phase). The
transition from $P_1$ phase to $P_3$ phase occurs at $T \simeq 0.82$. The high-temperature ordered phase ($P_1$) transition 
(from dynamically
disordered phase) occurs at $T \simeq 1.20$. These two transitions were reexamined by studying the temperature variations of the
derivatives of the dynamic order parameter and obtained the same transition temperatures. 

From the above discussion it is quite clear that if one studies the dynamic phase transitions by varying $h_{0x}$ only (keeping
$D=0.2$ and $h_{0z} = 1.0$ fixed), one would get two transitions for $h_{0x} =0.0$. Just above $h_{0x} =0.0$, i.e., starting from
$h_{0x}=0.1$ upto $h_{0x}=0.5$ one would get three transitions (and three ordered phases). Above this value, say $h_{0x} = 0.6$
one would get again two transitions (two ordered phases). If one continues further, it was
observed that the two transitions feature continues. 
The transition points (temperatures) are estimated by studying specific heat and the derivatives of the dynamic order parameter
components. These are not shown in figure, only the results (obtained from the peak positions of the specific heat
and positions of maximum or minimum of ${{dQ_{\alpha}}/{dT}}$) are taken.
It was
observed further that all the transition points shift towards lower temperatures for higher values of $h_{0x}$. The whole results
of these multiple transitions temperatures and plotted in $T-h_{0x}$ plane (fig. \ref{phase}) to get the multiple dynamic phase boundary. 

In the phase diagram (fig. \ref{phase}), the outermost boundary separtes the dynamically
$Y$-ordered phase $P_1:(Q_x = 0, Q_y \neq 0, Q_z =0)$ from the disordered phase $P_0: (Q_x = 0, Q_y = 0, Q_z = 0)$. 
The transition temperature decreases as the value of the field amplitude $h_{0x}$ increases.
For lower values of $h_{0x}$, the 
region bounded by the boundaries marked by the symbols circle and bullet is the ordered phase characterised as
$P_2: (Q_x \neq 0, Q_y = 0, Q_z \neq 0)$. This region is quite small (in area) but very clear and was observed very
carefully. 
The transition temperatures for both the boundaries (left and right) of this phase decrease as the field amplitude
$h_{0x}$ increases.  
The innermost (in the $T-h_{0x}$ plane) region, whose boundary is partly marked by circle (for higher values
of $h_{0x}$) and partly marked by bullet (for lower values of $h_{0x}$), represents the low-temperature phase characterised
by $P_3:(Q_x = 0, Q_y = 0, Q_z \neq 0)$. Here also, the transition temperature decreases as the value of the field 
amplitude $h_{0x}$ increases. 

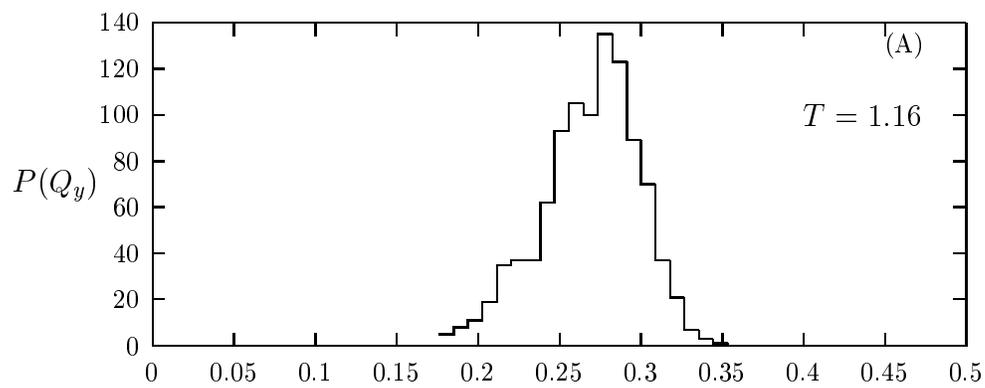
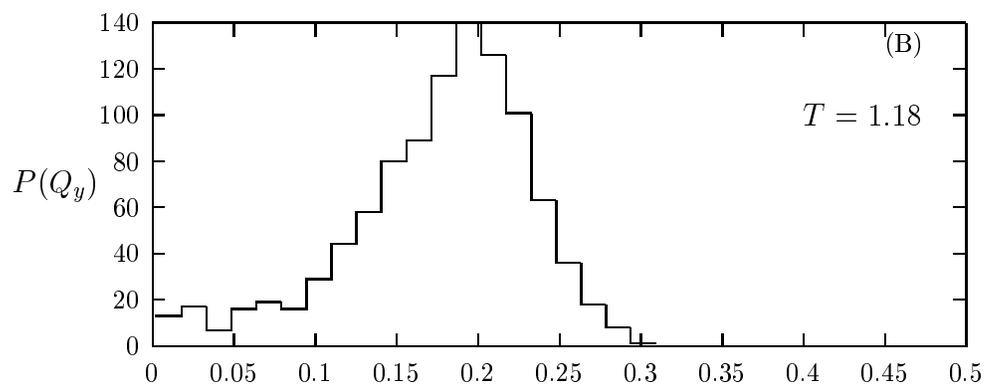
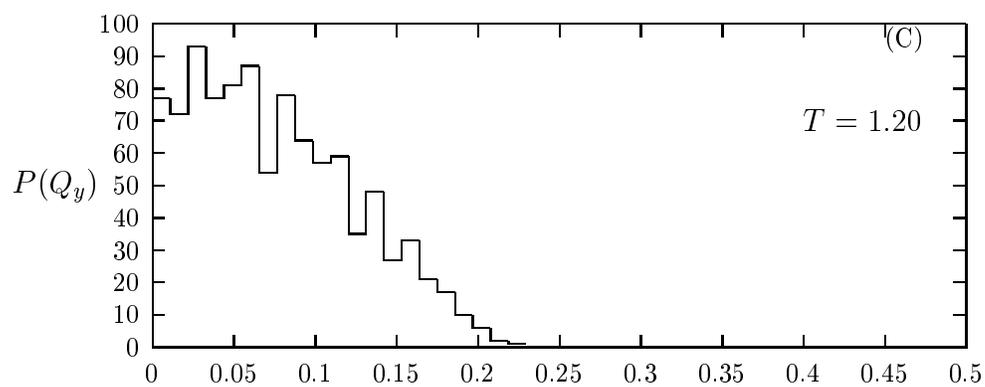
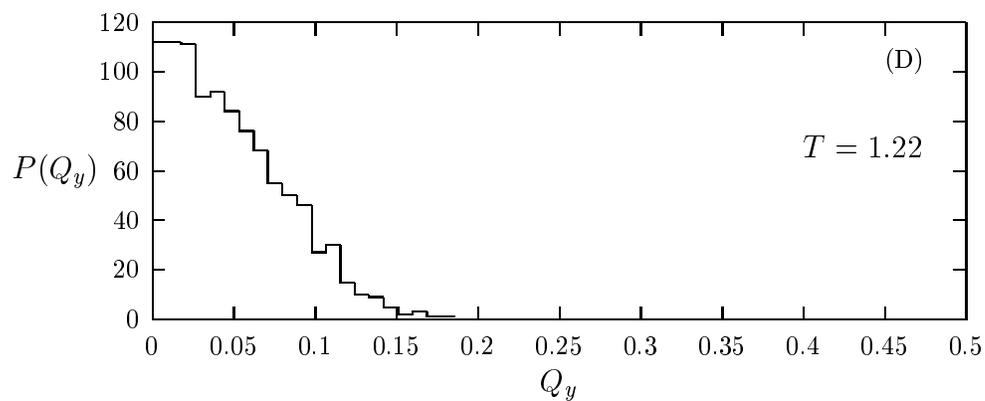
\begin{figure}
\caption {The unnormalised distributions of the magnitudes of dynamic order parameter component $Q_y$ for
different temperatures. Here, $f=0.02$, $D = 0.2$, $h_{0x} = 0.7$, $h_{0z} = 1.0$ and $f = 0.02$.}
\label{binder2}
\end{figure}

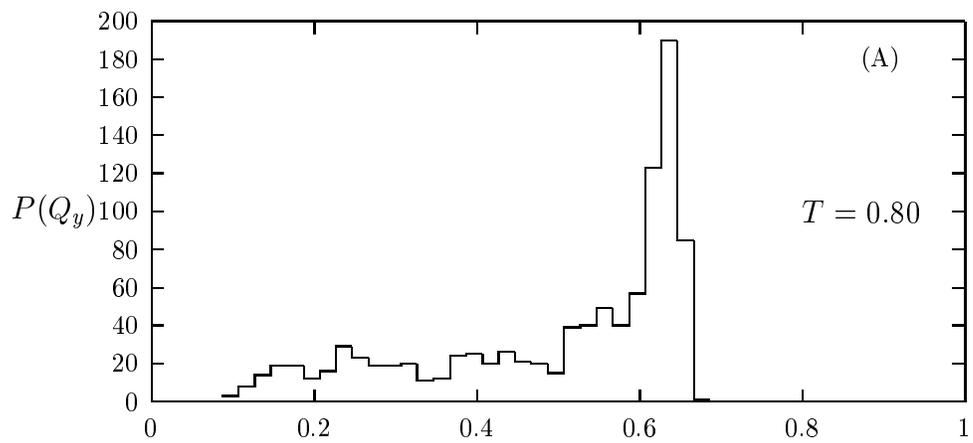
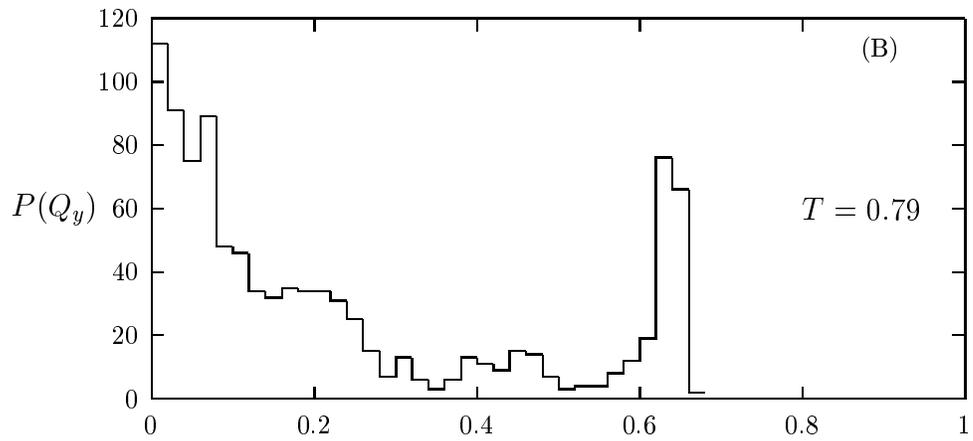
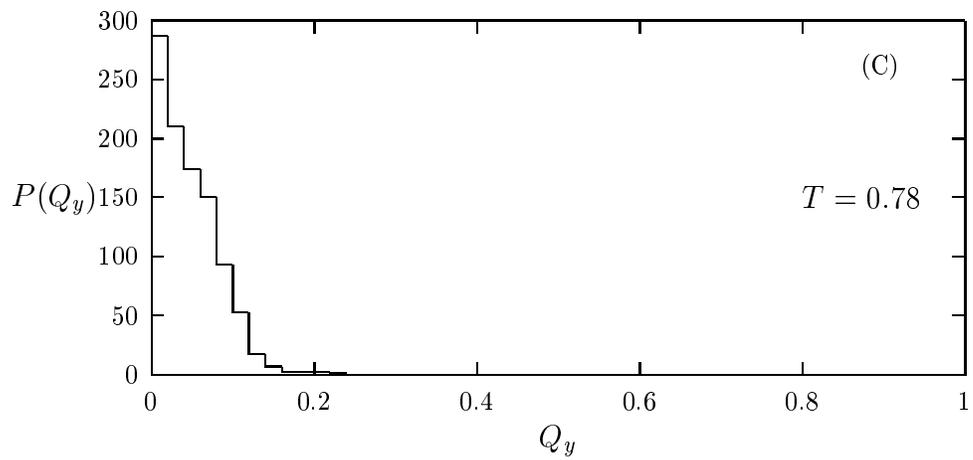
\begin{figure}
\caption {The unnormalised distributions of the magnitudes of the dynamic order parameter component $Q_y$
for different temperatures. Here, $D = 0.2$, $h_{0x} = 0.7$, $h_{0z} = 1.0$ and $f = 0.02$.}
\label{binder1}
\end{figure}

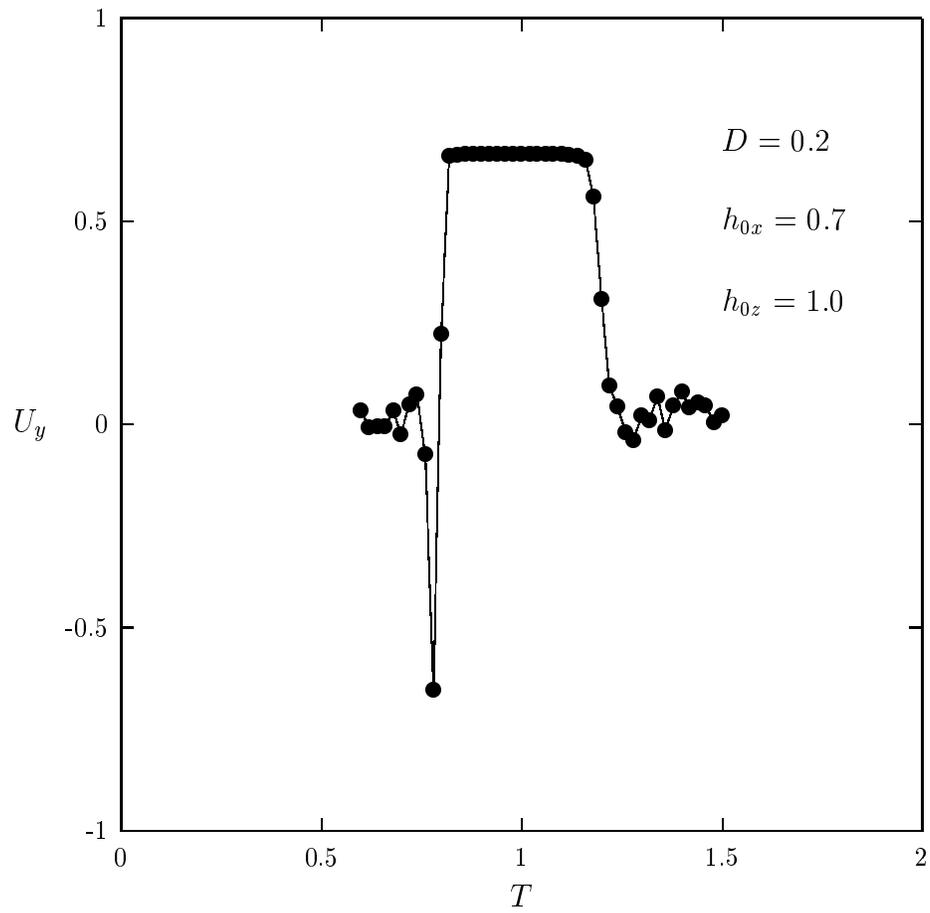
\begin{figure}
\caption {The temperature variation of the Binder cumulant ratio $U_y$ near the transition points.
Here, $D=0.2$, $h_{0x}=0.7$, $h_{0z}=1.0$ and $f=0.02$.}
\label{cum12}
\end{figure}

The question which should arise now,
what will be the nature (continuous/discontinuous) of these multiple dynamic transitions ? To get the answer of
this questions, the distributions of the magnitudes of the dynamic order parameter components are studied near the
transition points. On the phase boundary (Fig.\ref{phase}), a particular value of $h_{0x}(=0.7)$ is chosen. This
choice is not arbitrary. In this region, the phase boundaries are well separated to study comfortably
the temperature variations
of fourth-order Binder cumulant ratio \cite{binder} $U_y = 1 - <Q_y^4>/(3<Q_y^2>^2)$. 
Since the transition from $P_0$ to $P_1$
phase is indicated by $Q_y$ the distribution of the magnitude of $Q_y$ is studied near the transition point. The
histogram is shown in Fig.\ref{binder2} for four different temperatures around the transition temperature. This figure
shows that as one goes through the results of temperature from $T=1.16$ to that of higher temperature,
the peak of the distributions shifts towards $Q_y = 0$ continuously
and around $T=1.20$ this gets peaked near $Q_y \simeq 0$.
For slightly higher value of the temperature, say $T=1.22$, the distribution is peaked around $Q_y =0$, corresponding to
the disordered phase. 
This indicates the transition is continuous with $T_{c1} \simeq 1.20$ (for the transition from $P_0$ to $P_1$ phase) 
and one could guess it 
from the temperature variation of $Q_y$ and ${{dQ_y}/{dT}}$. 
This same technique was applied to study the nature of the transition
from $P_1$ to $P_3$ phase. The distribution of $Q_y$ is shown (in Fig.\ref{binder1}) for three different temperatures.
Here, the different (from the previous case just discussed above) senario was observed and
may be noted, that the value of $Q_y$ drops to zero from a nonzero value as the system gets cooled.  
At $T=0.80$ the distribution is singly peaked near $Q_y \simeq 0.63$. At slightly lower
temperature $T=0.79$, the distribution get doubly peaked. One additional peak appears near $Q_y \simeq 0$. This simultaneous
appearance of two peaks at very close to the transition temperature is a clear signature of discontinuous transition
with $T_{c} \simeq 0.79$ (for the transition from $P_1$ to $P_3$ phase). From this study, one can get the idea about the maximum error (0.01 here) 
in estimation of the transition
temperature.
As one lowers the temperatures, $T =0.78$, one gets only one peak of the distribution around $Q_y=0$ after the
transition. 
Here also, this could be guessed from the temperature variation of $Q_y$.
These studies indicate that the high-temperature transition is a continuous one and the low-temperature
transition is discontinuous type. These natures of the transitions were reexamined by studying the temperature variation
of Binder cumulant ratio $U_y$. 
For a continuous transition, $U_y$ should change monotonically from 0 to 2/3 as one tunes the system from the disordered
to the ordered phase. On the other hand, for a discontinuous transition, $U_y$ develops sharp minimum, whose location
corresponds to the transition point.
The results, obtained here, indeed show this (Fig.\ref{cum12}). Fig.\ref{cum12} shows, that as one tunes the system from
 high temperature, the Binder cumulant ratio $U_y$ first grows monotonically from 0 to 2/3 around $T_{c1} \simeq 1.20$
indicating that the high-temperature transition is continuous. As the temperature decreases further, it shows a very
sharp minimum around $T_{c} \simeq 0.80$ (for the transition from $P_1$ to $P_3$ phase) 
indicating the discontinuous nature of the low-temperature transition.   
These two (multi-histogram analysis and Binder cumulant ratio)
studies together confirm that the high-temperature transition is a continuous type and the low-temperature transition
is discontinuous type. 
Since, the evidence of the nature of the high-temperature transition is found to be of continuous type, it will be quite 
interesting to study the scaling behaviour (if any) with the estimations of critical exponents. But for this, one has to
estimate the transition temperatures more precisely which will require larger sizes of the system.   

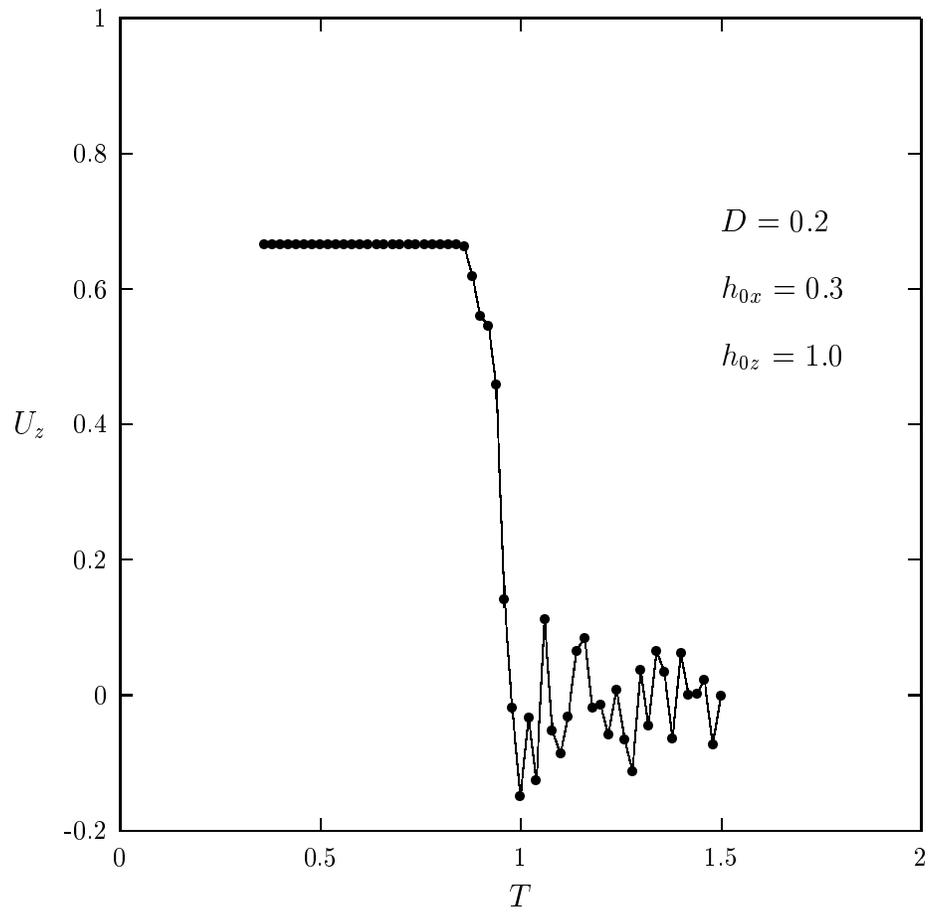
\begin{figure}
\caption {The temperature variation of the Binder cumulant ratio $U_z$ near the transition point.
Here, $D=0.2$, $h_{0x}=0.3$, $h_{0z}=1.0$ and $f=0.02$.}
\label{cum3}
\end{figure}

What will be the nature of the transition from $P_2$ to $P_3$ phase ? For lower values of
$h_{0x}$ (=0.3 chosen here)
one would be able to see this transition. From Fig.\ref{pt3}(a), this transition is indicated by the temperature
variation of $Q_z$. Since, this transition corresponds to very small change of $Q_z$, it is quite difficult to observe
any change of the distribution of $Q_z$ near the transition temperature. 
However, the temperature variation
of the Binder cumulant ratio $U_z = 1-<Q_z^4>/(3<Q_z^2>^2)$ indeed shows (see Fig.\ref{cum3}) 
smooth variation near the transition point and it
grows monotonically from 0 to 2/3 which indicates the transition is continuous with $T_{c3} \simeq 0.88$ (for the
transition from $P_2$ to $P_3$ phase). 
All these natures of the transitions are marked in the
phase diagram (Fig.\ref{phase}). 

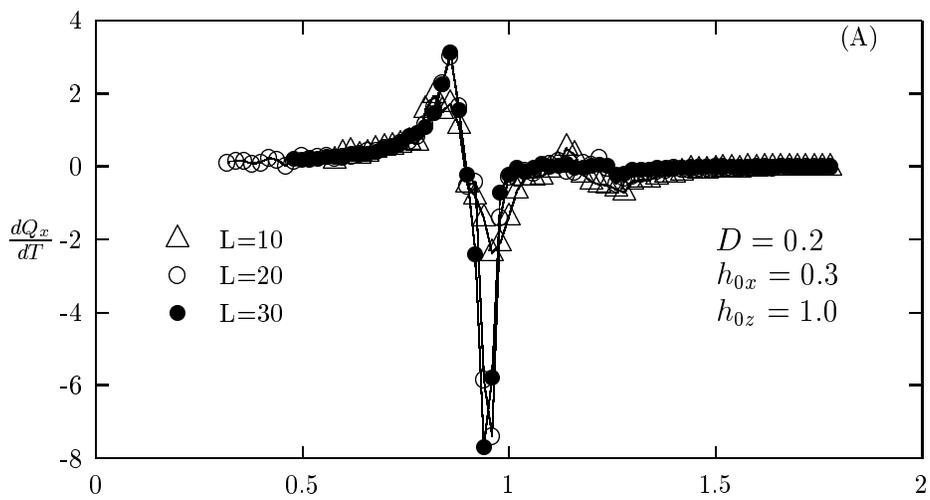
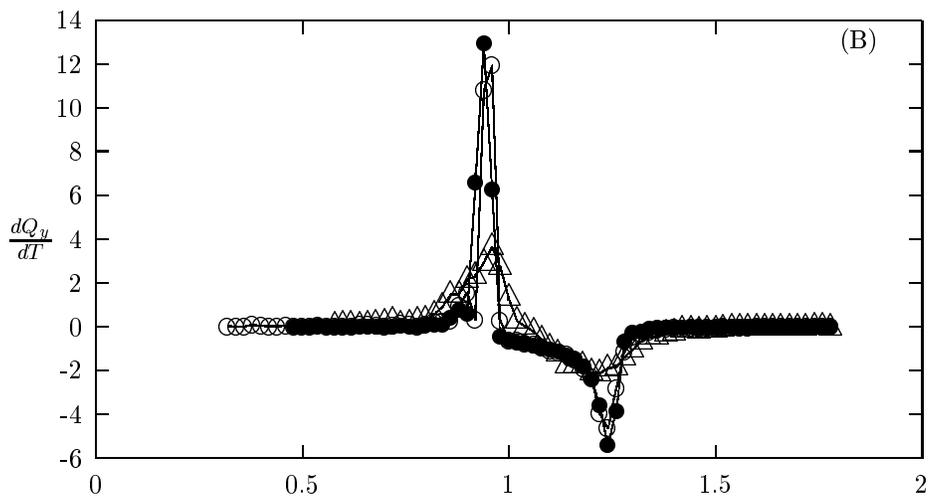
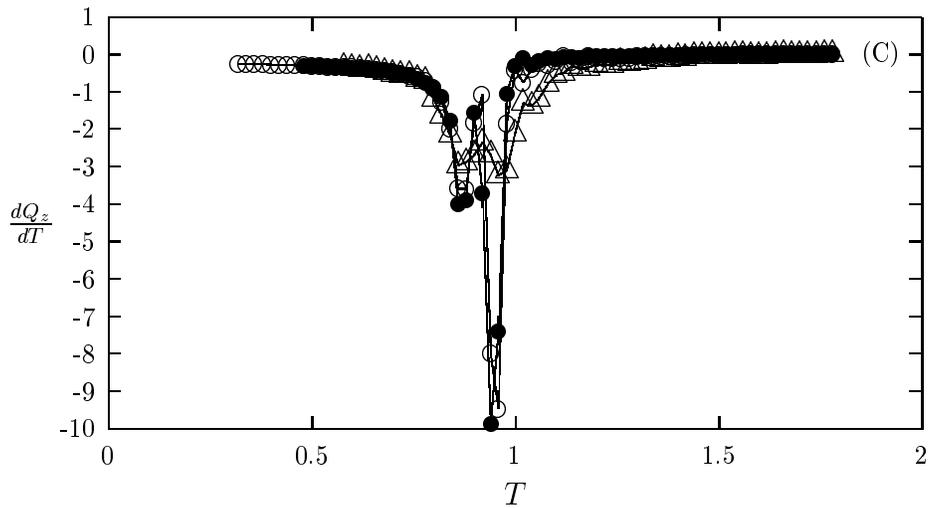
\begin{figure}
\caption {Temperature variations of derivatives of dynamic order parameter components for different system sizes
($L = 10, 20$ and 30) for $f=0.02$, $D=0.2$, $h_{0x} = 0.3$ and $h_{0z} = 1.0$.  
(A) ${{dQ_x}/{dT}}$ versus $T$.  (B) ${{dQ_y}/{dT}}$ versus $T$. 
(C) ${{dQ_z}/{dT}}$ versus $T$. Solid lines are guide to the eye.}
\label{fnt1}
\end{figure}

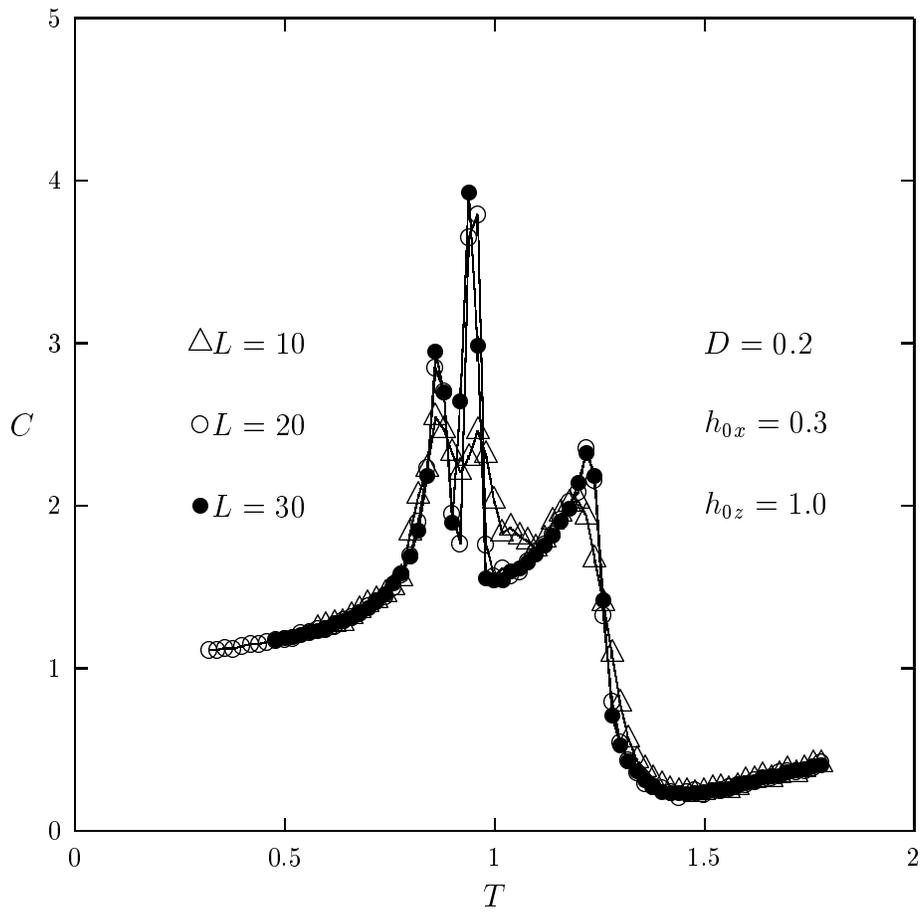
\begin{figure}
\caption {Temperature variations of dynamic specific heat for different system sizes ($L$ = 10, 20 and 30)
for $f=0.02$, $D$ = 0.2, $h_{0x}$ = 0.3 and $h_{0z}$ = 1.0. Solid lines are guide to the eye.}
\label{fnt3}
\end{figure}

The finite size study is also done to confirm that the observation is not an artifact of limited size of the system.
This study was done for a particular set ($f=0.02$, $D=0.2, h_{0x} =0.3, h_{0z} = 1.0)$ of values of the parameters.
Different statistical quantities were studied as functions of temperatures for different linear sizes ($L =
10, 20, 30$ here) of the system. This particular choice of the values of $D$, $h_{0x}$ and $h_{0z}$ is meaningful
in the sense that the three transitions phenomenon was observed clearly for this particular values of parameters
and the most important part of the phase boundary. 
Figure \ref{fnt1} shows the temperature variations of ${{dQ_{\alpha}}/{dT}}$ for $L = 10, 20$ and 30. From the
figure it is clear that both the sharpness and the height (depth) of maximum (minimum) indicating the transition points
increase as the system size increase. The dynamic specific heat is also studied as a function of temperature for different
$L$ and same set of other parameter values. This is shown in fig. \ref{fnt3}. This also indicates that the height of
the peaks (indicating the transitions) increases as the system size increases. The finite size study, done 
and briefly reported here, at
least may indicate that the multiple dynamic transitions mentioned above is not an artifact of finite size effect.

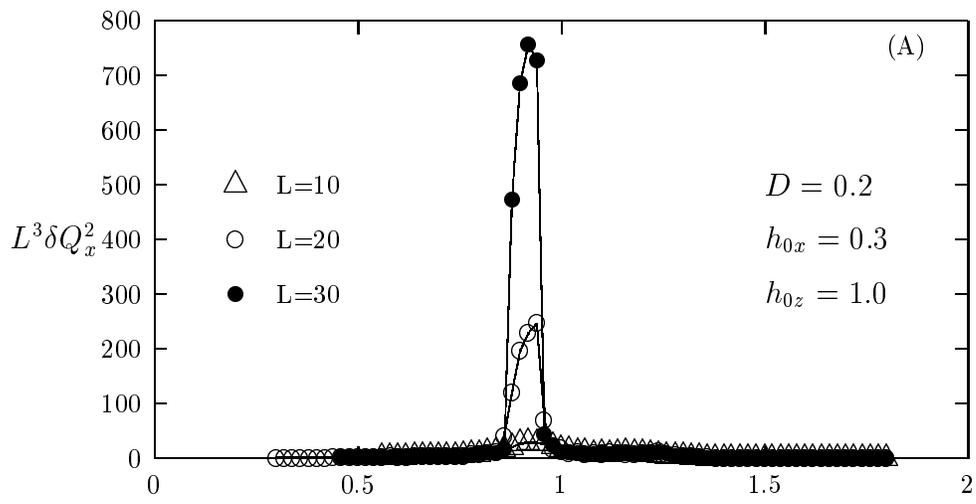
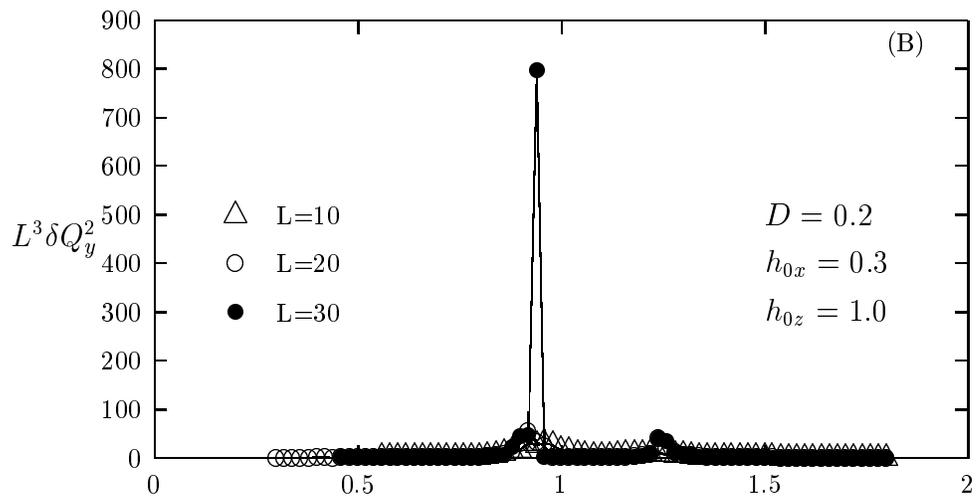
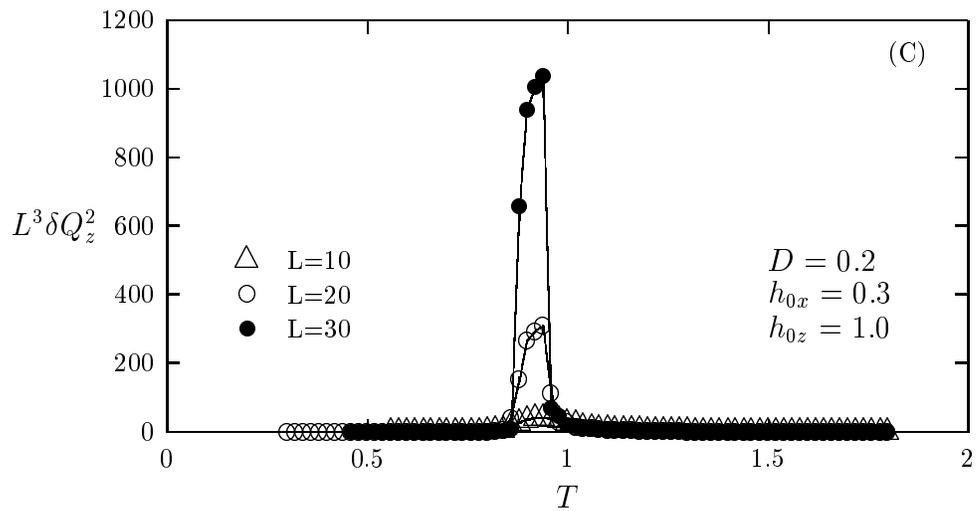
\begin{figure}
\caption {Temperature variations of variances of the dynamic order parameter components for different system sizes
($L$ = 10, 20 and 30) for $D$ = 0.2, $h_{0x}$ = 0.3 and $h_{0z}$ = 1.0. 
(A) $L^3 \delta Q_x^2$ versus $T$. (B) $L^3 \delta Q_y^2$ versus $T$. (C) $L^3 \delta Q_z^2$ versus $T$. Solid
lines are guide to the eye.}
\label{fnt2}
\end{figure}

The variances of the dynamic order
parameter components i.e., 
$L^3 {\rm Var}(Q_{\alpha})=L^3[<Q_{\alpha}^2>-<Q_{\alpha}>^2]=L^3\delta Q_{\alpha}^2$ 
are studied as function of temperature and for different
system sizes. The values of other parameters are $f=0.02$, $D=0.2$, $h_{0x} = 0.3$ and $h_{0z} = 1.0$. 
Figure \ref{fnt2}(a) shows temperature variation of
$L^3 {\rm Var}(Q_{x})=L^3[<Q_{x}^2>-<Q_{x}>^2]=L^3 \delta Q_x^2$ for $L = 10, 20$ and 30. It gets sharply 
peaked at around $T = T_{c2} \simeq 0.96$
(the transition temperature from $P_1$ phase to $P_2$ phase). 
The height of the peak increases for larger system sizes.
In fig. \ref{fnt2}(b) the $L^3 {\rm Var}(Q_y)$ is
plotted against temperature for different system sizes. It shows two peaks. The high-temperature peak (very short in height
but visible in this scale) is located at $T \simeq 1.22$. This corresponds to the transition from disordered ($P_0$) to first
Y-ordered ($P_1$) phase. The low-temperature peaks are positioned at around $T \simeq 0.96$ corresponding to the transition
from $P_1$ to $P_2$.
Here, also the heights of the peaks increases systematically as the system size increases.
Similar things are observed for $L^3 {\rm Var}(Q_z)$ and shown in fig. \ref{fnt2}(c). Here, an important thing
should be noted that this study indicates 
that their exists a diverging length scale at the dynamic transition
points \cite{rikpre}, for the multiple dynamic transitions in this system. It should also be noted that the transition
(from $P_2$ to $P_3$) temperature $T_{c3}$ cannot be resoved from $T_{c2}$ in the present study. It was already indicated
by the results shown in Fig.\ref{pt3}(a). The sizes of the errorbars of $Q_x$, $Q_y$ and $Q_z$ becomes maximum near $T_{c1}$ and $T_{c2}$.
However, the growth of the errorbars of $Q_z$ is not quite clear near $T_{c3}$. 

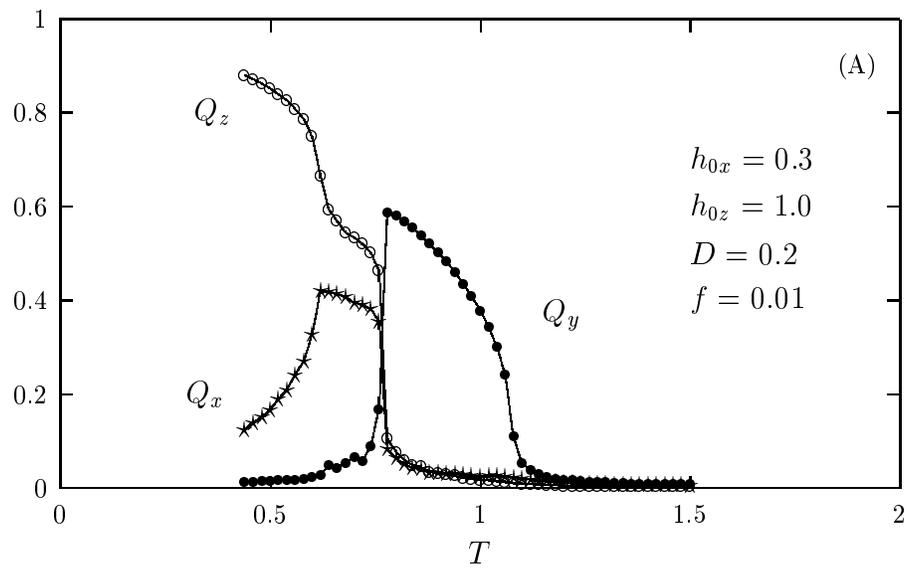
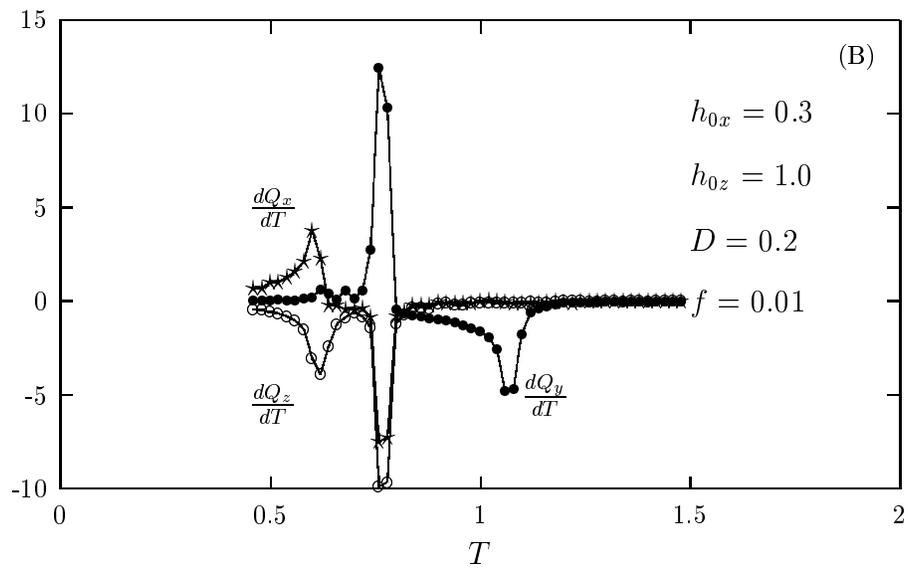
\begin{figure}
\caption {Temperature variations of the (A) magnitudes of dynamic order parameter components
(B) the derivatives of the order parameter components. Here, $D=0.2$, $h_{0x}=0.3$, $h_{0z} = 1.0$
and $f = 0.01$.}
\label{frq1}
\end{figure}

The frequency dependence of these multiple dynamic transitions are also studied very briefly. For a lower frequency
($f = 0.01$) of the polarised magnetic field, the multiple dynamic transition was studied with $D=0.2$, $h_{0x}=0.3$
$h_{0z}=1.0$. The temperature variations of the magnitudes of the dynamic order parameter components are plotted in
Fig.\ref{frq1}(a). This shows similar kind of temperature variations as that was obtained for $f=0.02$ ($D=0.2$,
 $h_{0x}=0.3$ and $h_{0z}=1.0$),
with considerable shifting of transition points towards lower temperatures. From the study of the temperature
variations of the derivatives of the dynamic order parameter components (shown in Fig.\ref{frq1}(b)) one gets
$T_{c1} \simeq 1.06$, $T_{c2} \simeq 0.76$ and $T_{c3} \simeq 0.62$. A similar study was done for higher values of 
$h_{0x} (=1.5)$ ($f=0.01$, $D=0.2$ and $h_{0z}=1.0$). The results are shown in Fig.\ref{frq2}. In this case
only two transitions were observed and the transition temperatures estimated are $T_{c1} \simeq 0.60$ and $T_{c3} \simeq 0.30$. 
From these studies, one can get the indication
that as the frequency decreases (or as one approaches the static limit) the transition points shifts towards lower temperatures
keeping the qualitative nature of the phase diagram (for $f=0.02$) unchanged. 
This is quite expected result, since as frequency decreases
the magnetisation components can follow the field in time 
and the dynamic transition disappears.  This is why the transition is 
called {\it truely dynamic} and discussed in details in earlier studies \cite{rmp} in the case of Ising model.

\begin{center} {\bf SUMMARY} \end{center}

The classical Heisenberg ferromagnet (uniaxially anisotropic) in presence of a time-varying polarised magnetic field
and in contact with a thermal bath is studied by Monte Carlo simulation using Metropolis dynamics. The magnetic field
is elliptically polarised and the field vector 
rotates on XZ - plane. For a very weakly anisotropic (uniaxial) system and particular
set of parameter values (amplitudes of field in x and z directions) if the system cools down it undergoes successive
phase transitions. For a range of values of field amplitude along x-direction, three dynamic phase transitions were
observed. Keeping the values of anisotropy strength and amplitude of field along z direction fixed, the phase transitions
were studied for different values of the field amplitude along the x direction. In a plane formed by the temperature
and the ratio of field amplitudes the phase diagram for the multiple dynamic phase transition was drawn. 
The nature (continuous/discontinuous) of the transitions are investigated by the temperature variations of the
Binder cumulant ratio and the distribution of order parameter near the transition points.
The
finite size study was done to check that the transitions are not artifacts of limited system size. The variances of
dynamic order parameter components are studied as a function of temperature taking system size as parameter. This
particular study indicates the existence of a diverging length scale at the transition points. 
The choice of the parameter values,
in the present study, are obtained from various trials. In other parameter range the multiple transitions
are also possible to observe, however this particular choice gives quite distinct results of 
the observed {\it multicritical behaviour} \cite{multicritical}. Since, the transition across the
outermost boundary of the phase diagram is found to be continuous, it will be much interesting to study the
scaling behaviour and to estimate the critical exponents. 
The frequency variations of these multiple dynamic transitions are also studied very briefly. This shows that the
qualitative nature of the phase diagram remains unchanged however the transition temperatures are lowered by lowering
the frequency of the polarised magnetic field. This is, of course, an expected result 
experienced from the earlier studies \cite{rmp}, particularly, in 
the case of the dynamic transition in Ising 
model.

The dynamic transition in Ising ferromagnet can be explained simply by spin reversal and nucleation \cite{nucl}.
The multiple dynamic transition occurs in Heisenberg ferromagnet possibly due to the coherent 
spin rotation \cite{uli}. To get
the clear idea about it one has to study also the dynamic configurations of spins in details. The study of the
relaxation dynamics of the system in the low anisotropic limit may help to analyse few results.

The alternative methods of studying this multiple dynamic phase transitions in anisotropic Heisenberg ferromagnet
driven by polarised magnetic field are: (i) to study Landau-Lifshitz-Gilbert equation of motion \cite{hinzke} with
Langevin dynamics, (ii) spin wave analysis of anisotropic Heisenberg ferromagnet in presence of polarised magnetic
field.

\begin{center} {\bf ACKNOWLEDGMENTS}\end{center}

The library facility provided by Saha Institute of Nuclear Physics, Calcutta, India, is
gratefully acknowledged. 

\vskip 1 cm

\newpage

\begin{figure}
\caption {The temperature variations of different dynamical quantities for
 $f=0.02$, $D=0.2$, $h_{0x} =0.3$ and $h_{0z} = 1.0$. 
(A) The magnitudes of components of dynamic order parameter $Q_x$, $Q_y$ and $Q_z$. Vertical lines are errorbars.
Note that the sizes of the errorbars become maximum near the transition points.
(B) derivatives of dynamic order parameters, (C) the dynamic energy and the (D) dynamic specific heat.
Solid lines are guide to the eye. The vertical arrows in (C) and (D) indicates the transition points.
The maximum error in estimating the transition temperature is 0.01.} 
\label{pt3}
\end{figure}

\begin{figure}
\caption {The temperature variations of (A) dynamic specific heat and 
(B) magnitudes of the order parameter components. For $f=0.02$, $D=0.2$, $h_{0x} =0.0$ and $h_{0z} = 1.0$.
Vertical lines are errorbars. Note that the sizes of the errorbars becomes maximum near the transition points.
Solid line in (A) is guide to the eye.}
\label{pt0}
\end{figure}

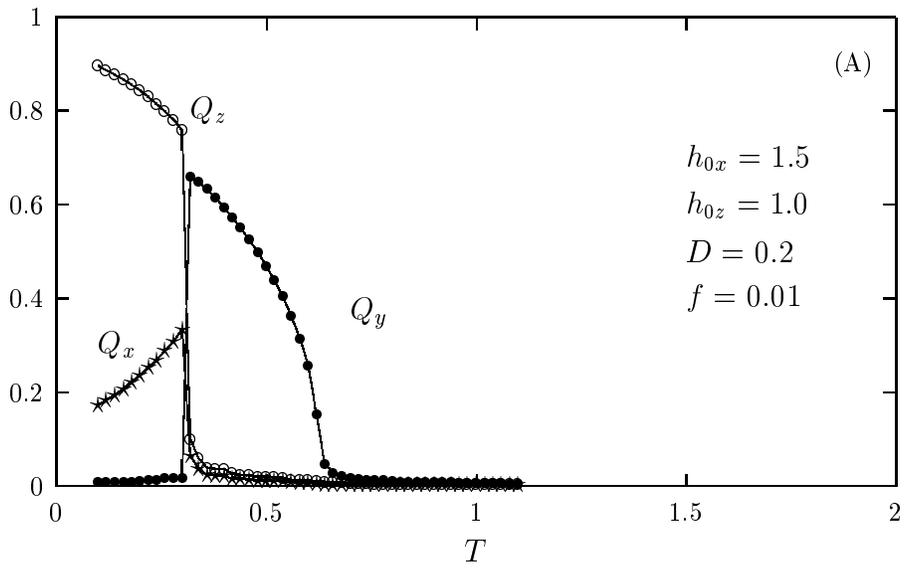
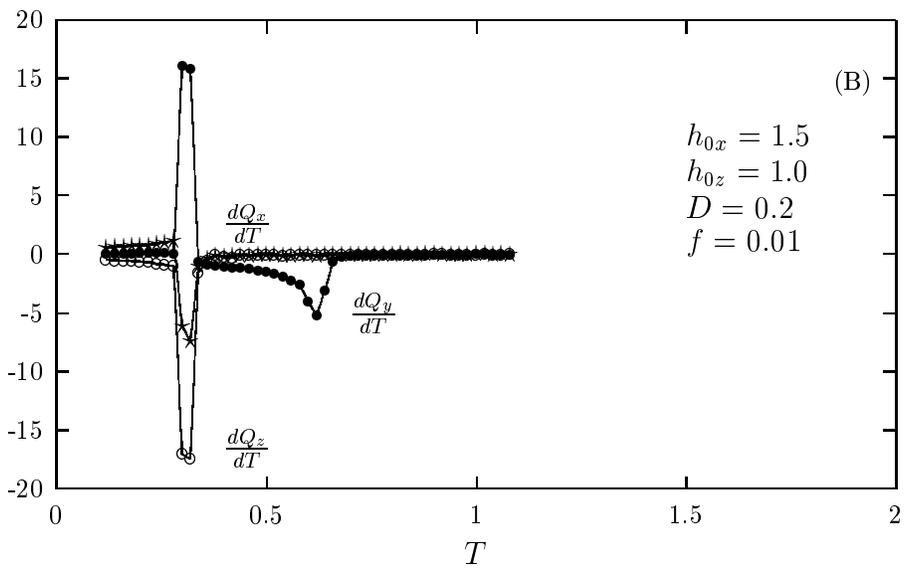
\begin{figure}
\caption {Temperature variations of the (A) magnitudes of dynamic order parameter components
(B) the derivatives of the order parameter components. Here, $D=0.2$, $h_{0x}=1.5$, $h_{0z} = 1.0$
and $f = 0.01$.}
\label{frq2}
\end{figure}
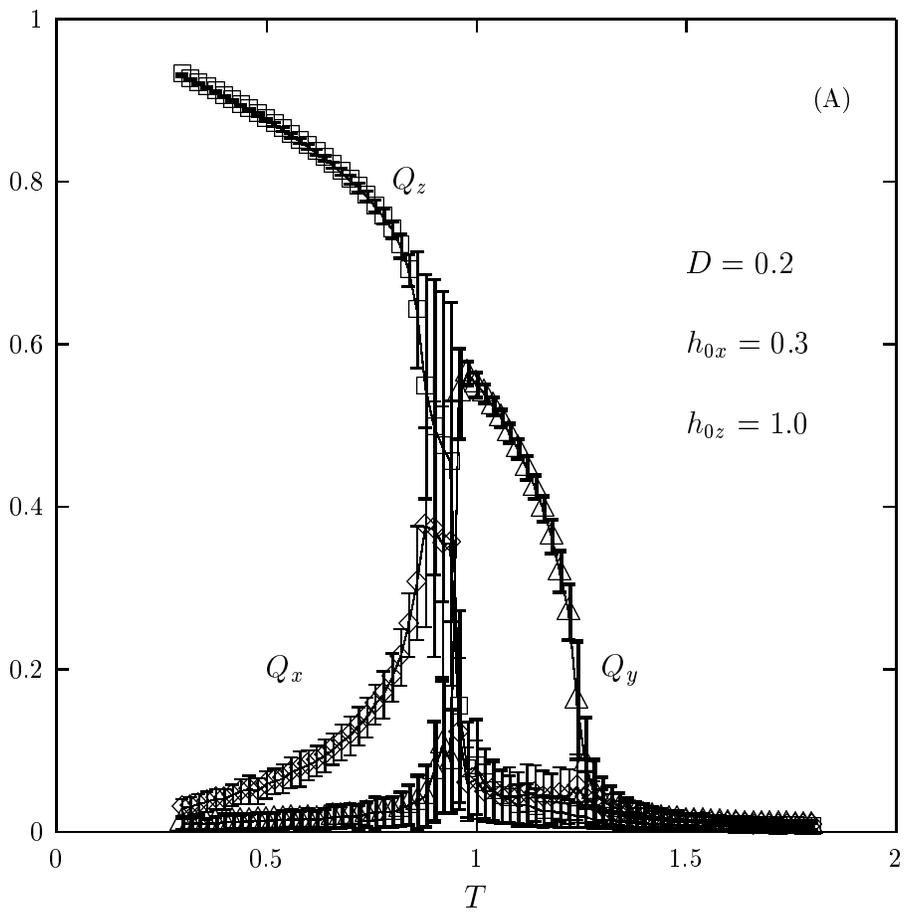
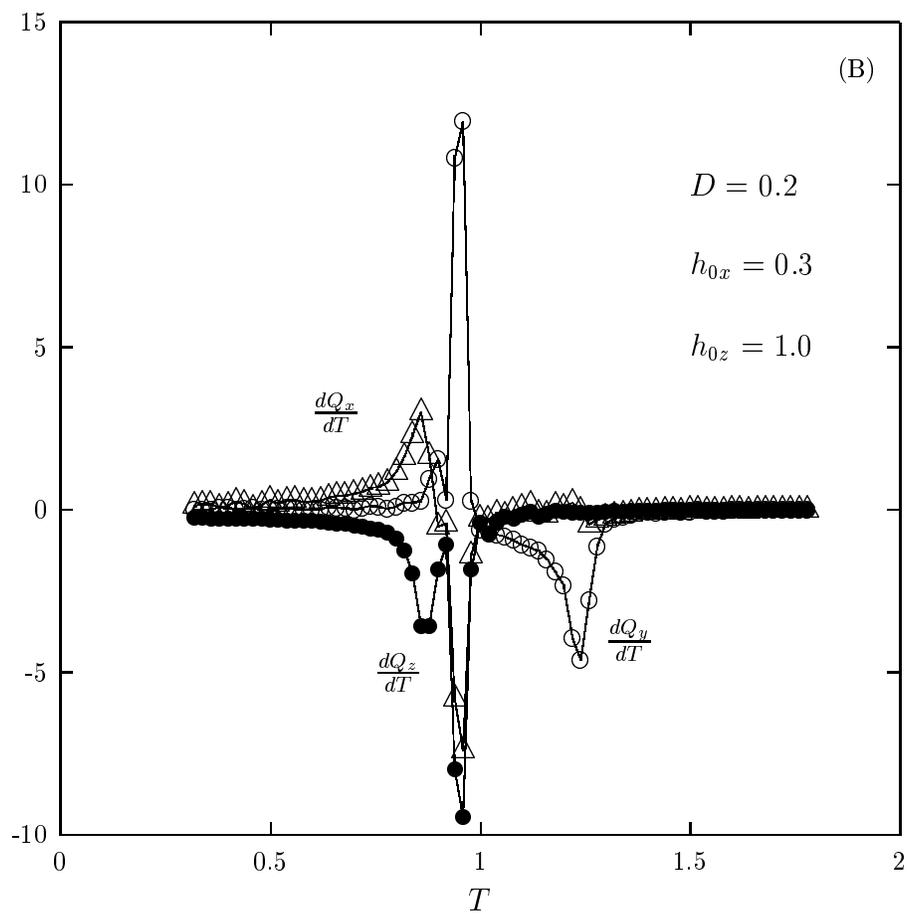
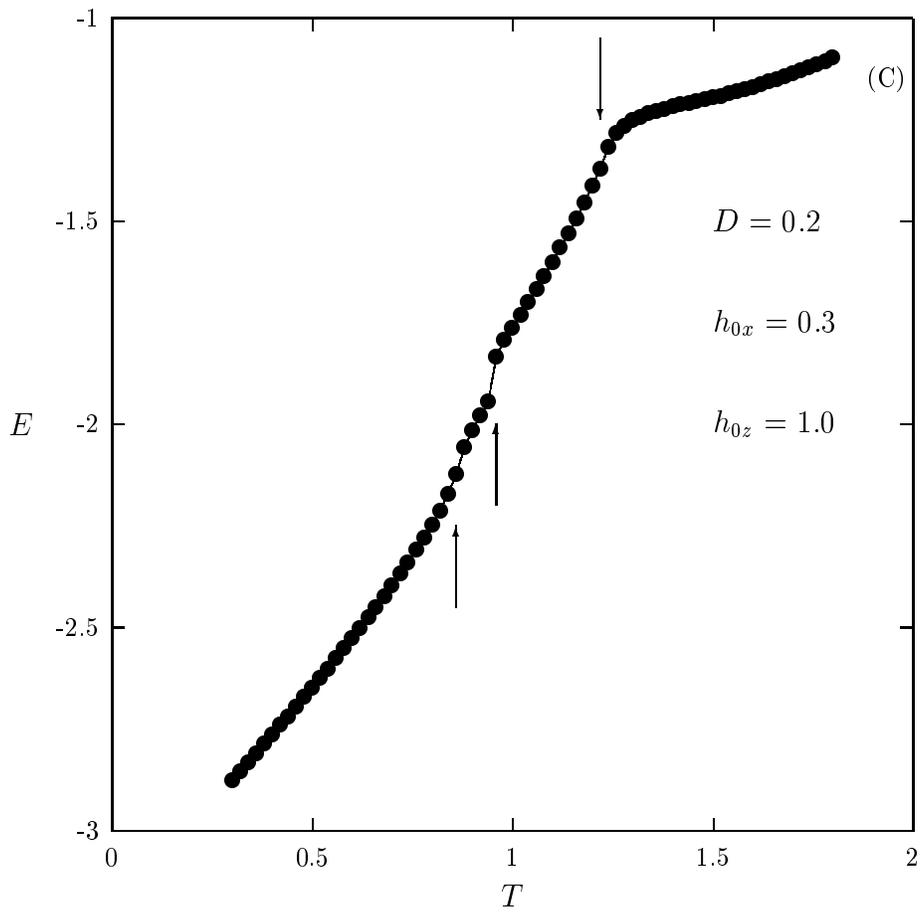
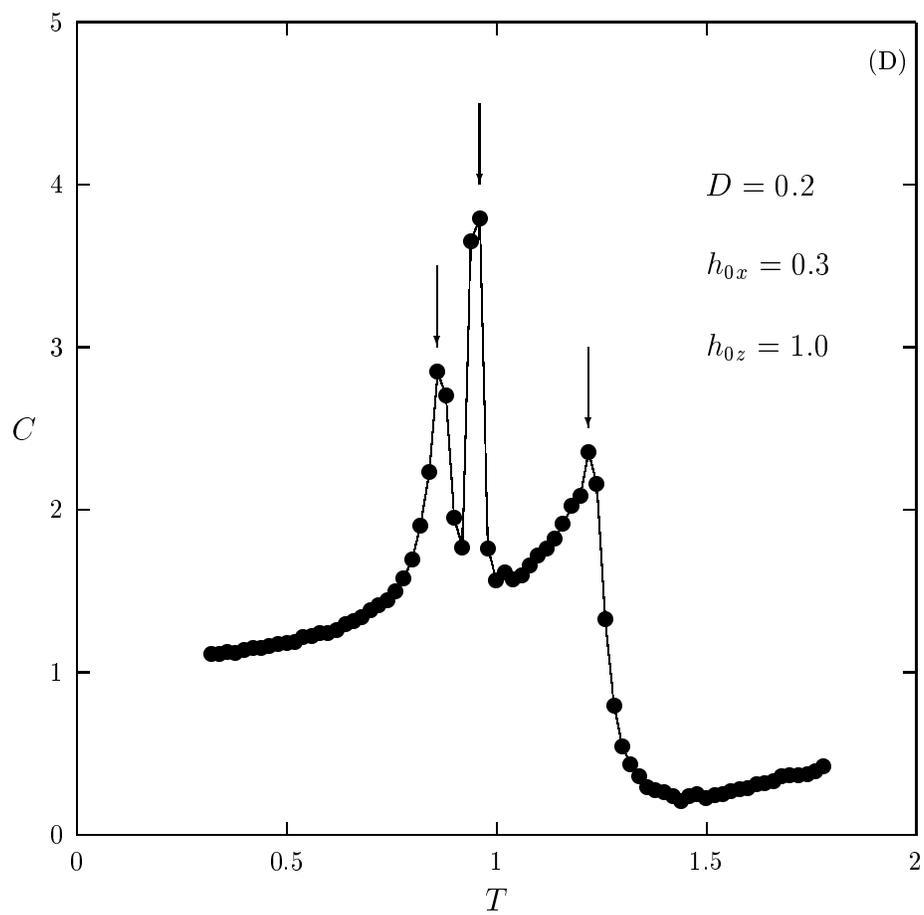
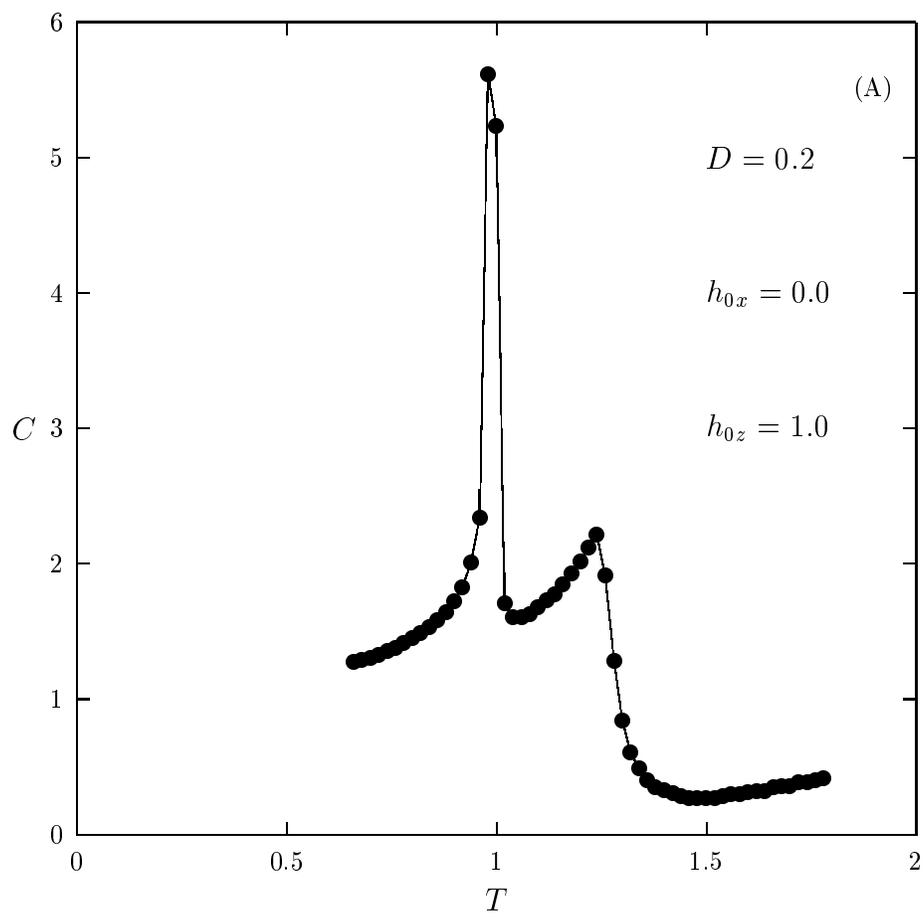
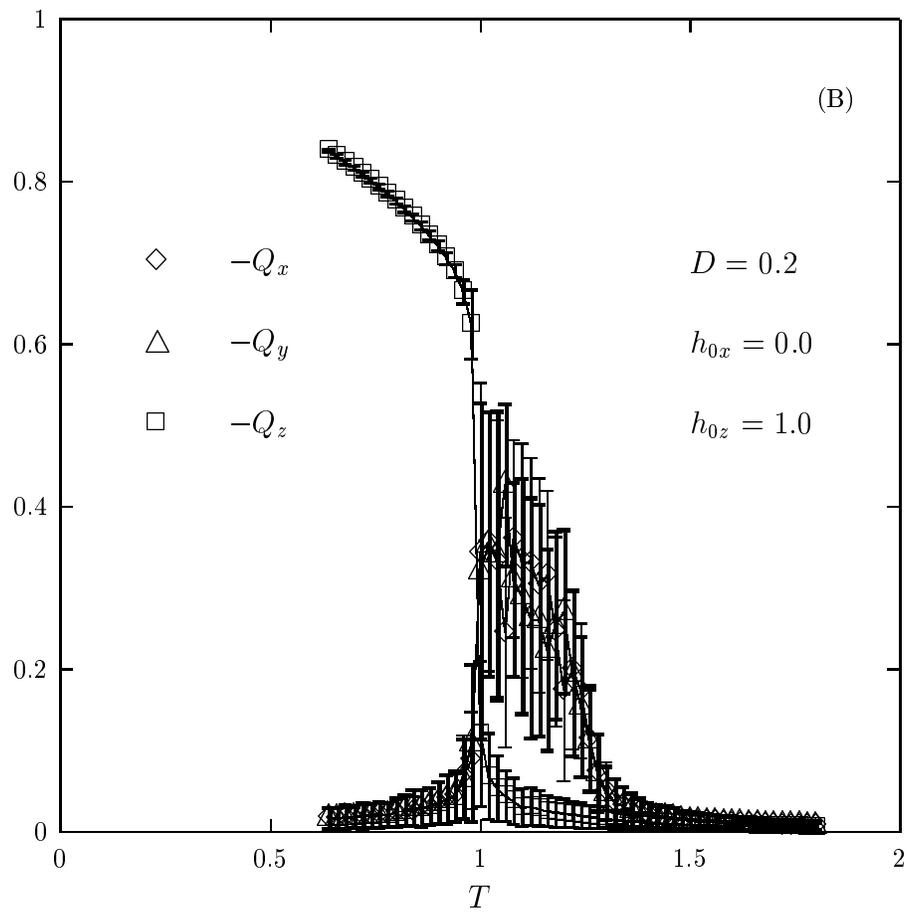
\end{document}